# Interrogation laser for a strontium lattice clock

T. Legero, Ch. Lisdat, J.S.R. Vellore Winfred, H. Schnatz, G. Grosche, F. Riehle, and U. Sterr

*Abstract*— We report on the setup and characterization of a 698 nm master-slave diode laser system to probe the $^1S_0 - {}^3P_0$ clock transition of strontium atoms confined in a one-dimensional optical lattice. A linewidth in the order of around 100 Hz of the laser system has been measured with respect to an ultrastable 657 nm diode laser with 1 Hz linewidth [1] using a femtosecond fiber comb as transfer oscillator [2,3]. The laser has been used to measure the magnetically induced $^1S_0 - {}^3P_0$ clock transition in $^{88}$Sr where a linewidth of 165 Hz has been observed. The transfer oscillator method provides a virtual beat signal between the two diode lasers that has been used to phase lock the 698 nm laser to the 1 Hz linewidth laser at 657 nm, transferring its stability to the 698 nm laser system.

## I. INTRODUCTION

Optical lattice clocks with strontium have reached relative uncertainties below $10^{-15}$ [4-6]. Their stability is ultimately limited by the quantum projection noise which for uncorrelated $10^6$ atoms and a Fourier-limited linewidth of 1 Hz would result in an Allan deviation of the relative frequency fluctuations of $\sigma_y(\tau) \approx 10^{-18} \cdot \tau^{-1/2}$. However, to reach this short term stability an even better stability of the clock laser is needed. Therefore an improved optical source which is phase stable during the required interrogation time of a few seconds represents a key technology for optical frequency metrology [7-9]. Until a few years ago, the investigation of the instability of a laser system was only possible by comparison with a similar system operating at the same wavelength. With the development of the frequency comb technology [10-13] it is now possible to measure the instability and linewidth of an ultra narrow laser by comparison with another laser system operating at a different wavelength [2,3]. In addition the use of frequency combs can be combined with a phase locking technique to stabilize a laser system with large intrinsic linewidth to a reference laser with ultra-narrow linewidth [3]. In this way the stability of a reference laser can be transferred to any other laser system. Thus the task to design a clock laser system for a certain frequency standard can be split into two parts: the design of an ultra-stable reference laser at a wavelength which allows for the ultimate stability and reliability, and the transfer of this stability to a less sophisticated laser system operating at the desired wavelength.

This paper is organized as follows: after a short description of a master-slave diode laser system designed for probing the 698 nm clock transition of strontium atoms confined in a 1D optical lattice (section II), we discuss the characterization of this laser system by comparing its frequency with that of a 1 Hz linewidth laser at 657 nm used in a Ca clock [1] (section III). In section IV we present results on the magnetically induced spectroscopy of the $^1S_0 - {}^3P_0$ clock transition of the bosonic strontium isotope $^{88}$Sr. The virtual beat signal derived with the transfer oscillator method is used to phase-lock the 698 nm laser system to the 1 Hz linewidth laser. We discuss this phase locking method in section V.

## II. SETUP OF THE CLOCK-LASER SYSTEM

For probing the $^1S_0 - {}^3P_0$ strontium clock transition we have set up a 698 nm cavity stabilized master-slave diode laser system as shown in Fig 1. The cavity and laser setup follows the design of the 657 nm laser [1]. For reducing seismic noise the cavity is placed on a vibration isolation platform inside an

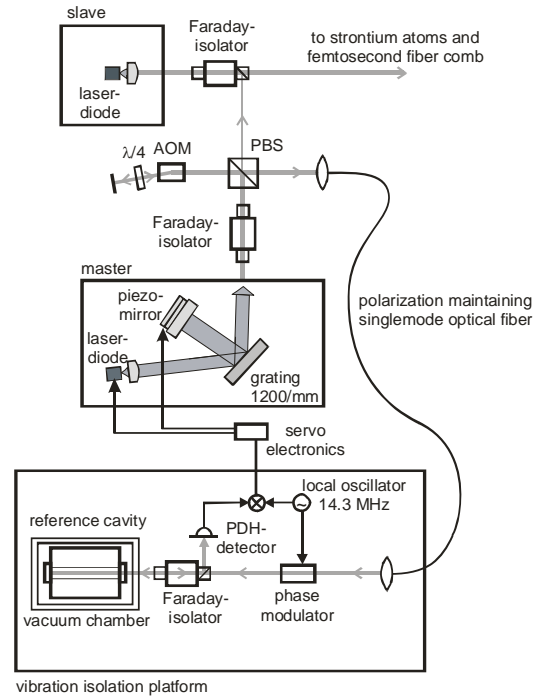

Fig. 1: Setup of the master-slave diode laser system for probing the strontium clock transition.

Manuscript received June 13, 2008. The authors gratefully acknowledge the support by the Deutsche Forschungsgemeinschaft under SFB 407, by ESA, DLR, and the Centre for Quantum Engineering and Space-Time Research (QUEST).

T. Legero is with the Physikalisch-Technische Bundesanstalt, Braunschweig, 38116 Braunschweig, Germany (phone: ++49-531-592-4306; fax: ++49-531-592-4305; e-mail: Thomas.Legero@ptb.de).

acoustic isolation box. Because of limited space and to avoid disturbing the isolation system both lasers are placed on a separate 60 cm × 90 cm breadboard. A short polarization maintaining single mode optical fiber of 1.5 m length links the master laser with the cavity setup. As was shown in the similar setup of the 657 nm laser the vibrational and thermal fiber noise is not limiting the master laser stability at the level of 1 Hz linewidth [1].

The master laser is an extended cavity diode laser (ECDL) in Littman configuration operated at a diode temperature of 44 °C with an output power of 4 mW. Its output frequency is locked to a high finesse optical cavity by the Pound-Drever-Hall (PDH) [14] stabilization technique. An acousto-optic modulator (AOM) introduces a frequency offset for tuning the laser. The cavity is made of a 100 mm long spacer made of ultra low expansion glass (ULE). The optical axis of the cavity is oriented horizontally. A cavity finesse of 330 000 has been measured, corresponding to a linewidth of 4.5 kHz. The cavity is mounted in a temperature stabilized vacuum chamber at a residual pressure of $10^{-7}$ mbar.

To minimize the sensitivity against vertical vibrations, the cavity is supported at four points near to its horizontal symmetry plane [15]. The support design is slightly different from that used in the 657 nm laser. In place of four holes drilled in the cylindrical cavity spacer, we glued small drilled invar plates on the spacer surface and use Viton cylinders to support the cavity. A sensitivity of about 140 kHz/ms$^{-2}$ to vertical vibrations was measured, which is far bigger than calculated from a finite element analysis. A difference of 1 mm from the optimal support point positions would result in a residual sensitivity of less than 8 kHz/ms$^{-2}$. Thus the observed sensitivity can not be explained by tolerances of the support point positions. We therefore attribute the discrepancy to the spacer touching the surrounding heat shield. This will be corrected in the near future.

The injection-locked slave laser delivers an output power of 23 mW. The laser remains injection-locked over several days without manual resetting and adjustment. Its light is sent to the strontium atoms and to the femtosecond fiber-laser comb by two optical fibers with provisions to cancel noise acting on the fiber length [16].

### III. CHARACTERIZATION OF THE PROBE LASER SYSTEM

A commercial femtosecond fiber-laser comb was used to characterize the laser system. We used the femtosecond laser as transfer oscillator [2,3] as shown in Fig. 2 to compare the optical frequencies $\nu_{Ca}$ and $\nu_{Sr}$ of the 657 nm Ca laser and the 698 nm Sr laser. For each laser the beat signal $\Delta_{Ca}$ and $\Delta_{Sr}$ with the neighboring comb line is detected. These beats are generated with the frequency-doubled output of the comb. Thus the frequencies of the comb lines are given by the product of the integer mode number m and the repetition rate $f_{rep}$ plus twice the carrier envelope offset frequency $2\nu_{CEO}$, which is measured with a $f - 2f$ interferometer using the original fs-laser output. Each beat signal is pre-processed by a tracking oscillator and digitally divided by two. Then $\nu_{CEO}$ is removed from both signals by multiplying them with $\nu_{CEO}$ in a mixer and selecting the sum frequency. The signal frequency in the calcium branch is multiplied by $m_1/m_2$ using a direct digital synthesizer (DDS). After subtracting the frequencies from each other one gets a signal, that corresponds to a virtual beat of the two optical frequencies $\nu_{Sr}/2$ and $\nu_{Ca} \cdot m_1/2m_2$. This signal follows the frequency fluctuations of both lasers with a bandwidth of several tens of kilohertz, limited by the bandwidth of various phase-locked loops used to track the intermediate frequencies. This virtual beat has a frequency of around 26 MHz. Since it is independent of $f_{rep}$ and $\nu_{CEO}$ it is not degraded by fluctuations of these two quantities within the tracking bandwidth.

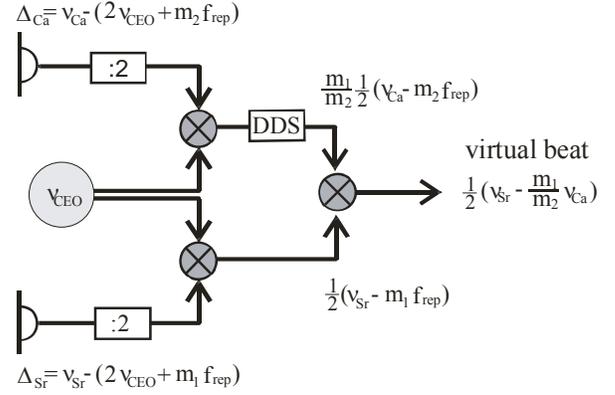

Fig. 2: Schematic of the rf electronics for generating a virtual beat signal between the strontium and the calcium laser. The fs-comb acts as a transfer oscillator. Thus the virtual beat is independent of fluctuations of the repetition rate $f_{rep}$ and the carrier envelope offset frequency $\nu_{CEO}$ of the femtosecond fiber comb.

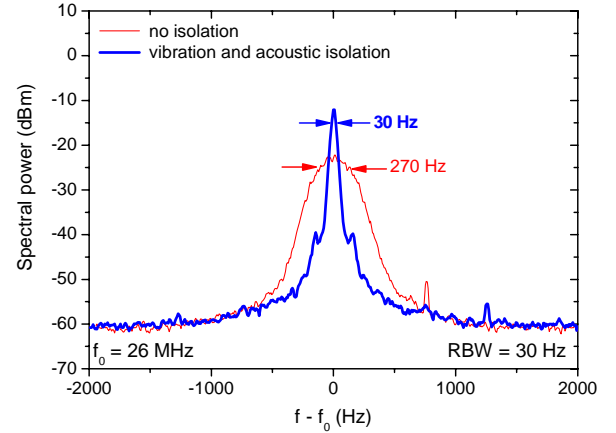

Fig. 3: Frequency spectrum of the virtual beat between the 698 nm Sr laser and the Ca reference laser. The spectrum obtained without any isolation (270 Hz linewidth) was averaged over 10 sweeps of 0.135 s. The spectrum measured with a passive vibration isolation table and acoustic isolation (30 Hz linewidth) was averaged over 100 sweeps.

Monitoring the beat signal with a spectrum analyzer provides an excellent tool for a real time analysis of the Sr clock laser. We use this setup to characterize and improve the 698 nm laser with respect to the 657 nm reference laser. Fig 3 shows the measured power spectral density of the virtual beat. Without any isolation of the reference cavity from seismic vibrations and acoustic noise the virtual beat note shows a linewidth of around 270 Hz. Using a vibration isolation platform and an additional box for acoustic noise reduction narrows the linewidth by a factor of 10 resulting in 30 Hz linewidth. The virtual beat is observed at half the frequency of the Sr laser. Depending on the spectrum of frequency fluctuations, the linewidth of the laser is between twice and four times the measured virtual beat linewidth [17], i.e. from this measurement we deduce a laser linewidth between 60 Hz and 120 Hz.

## IV. SPECTROSCOPY OF THE STRONTIUM CLOCK TRANSITION

Using the 698 nm laser system with the cavity on the vibration isolation table but without the acoustic isolation box we investigate the single photon excitation of the $^1S_0 - {}^3P_0$ clock transition in bosonic $^{88}$Sr. The strontium atoms are confined in a 1D optical lattice. To load the atoms into the optical lattice we cool the strontium atoms to a few microkelvin using a two-stage cooling process.

In the first cooling stage, atoms are captured from a Zeeman-slowed atomic beam and cooled to 2 mK in a magneto-optical trap (MOT) operating on the broad $^1S_0 - {}^1P_1$ transition at 461 nm [18,19]. This MOT works with a magnetic field gradient of 7.4 mT/cm, a $1/e^2$ laser beam diameter of 10 mm and a total laser intensity of 21 mW/cm$^2$. The cooling laser is detuned 54 MHz below the $^1S_0 - {}^1P_1$ transition frequency. After 200 ms $3 \cdot 10^7$ atoms are trapped in the MOT. For further cooling, a MOT working at the spin-forbidden $^1S_0 - {}^3P_1$ transition at 689 nm with a $1/e^2$ laser beam diameter of 5.2 mm is employed [18,19]. To cover the Doppler shift of the atoms from the first cooling stage and to compensate the limited velocity capture range of the 689 nm MOT the laser spectrum is broadened by modulating the laser frequency at 50 kHz with a peak to peak frequency excursion of 3 MHz. For this phase of the 689 nm MOT a magnetic field gradient of about 0.7 mT/cm, a total intensity of 33 mW/cm$^2$ and a detuning of 1.6 MHz below the $^1S_0 - {}^3P_1$ transition is used. Within a 70 ms long broadband cooling interval the atoms were cooled down to 15 µK. Finally the frequency modulation is switched off and the cooling laser is operated at a single frequency detuned 400 kHz below the $^1S_0 - {}^3P_1$ transition. With an intensity of 440 µW/cm$^2$ and a 70 ms long cooling interval this process leads to $8 \cdot 10^6$ atoms at a temperature of 3 µK.

During the whole cooling process the atomic cloud is superimposed with the horizontally oriented 1D optical lattice operated at 813 nm. At his wavelength the light shift of the $^1S_0$ and $^3P_0$ states cancel and the clock transition frequency becomes independent of the laser intensity [20,21]. As shown in Fig. 4 the 1.1 W output beam of the Ti:sapphire lattice laser is coupled into a polarization maintaining optical fibre and passes through polarization optics before being focused on the center of the atom cloud. The beam is linearly polarized with its polarization oriented perpendicular to gravity. A dichroic mirror is used to retro-reflect the 813 nm laser beam and hence establish the 1D optical lattice. With a beam radius of 30 µm and a power of 600 mW a trap depth of 120 µK is realized. After switching off the 689 nm MOT up to $10^6$ atoms at 3 µK are trapped in the lattice. This corresponds to a transfer efficiency from the 461 nm MOT into the lattice of up to 3%. A shorter loading time of the 461 nm MOT results in less atoms in the lattice. For the spectroscopy of the $^1S_0 - {}^3P_0$ transition we choose a loading time of 13.5 ms resulting in $1.2 \cdot 10^5$ lattice trapped atoms to avoid collision broadening of the clock transition. The light of the 698 nm slave laser is superimposed with the 1D optical lattice. The beam has a waist radius of 40 µm.

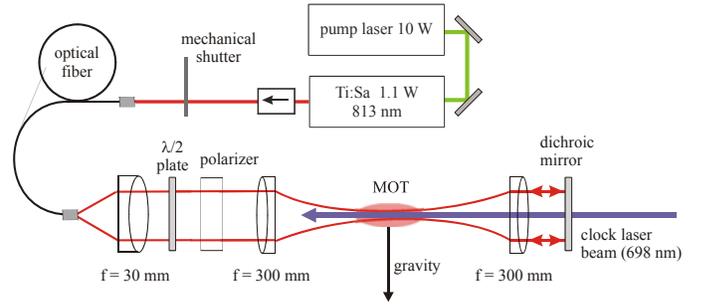

Fig. 4: Setup of the 1D optical lattice at 813 nm. The 1 D lattice is directed perpendicular to gravity and to the axis of the MOT coils.

To enable the $^1S_0 - {}^3P_0$ clock transition in bosonic $^{88}$Sr, which is forbidden for any single photon transition, we follow the proposal by Taichenachev et al. [22] and apply a dc magnetic field for mixing a small and controllable fraction of the nearby $^3P_1$ state to the $^3P_0$ state. This method has been successfully employed by Z. W. Barber et al. with neutral $^{174}$Yb [23] and by X. Baillard et al. with $^{88}$Sr [24]. The magnetic field is oriented parallel to the linear polarization of the interrogation laser beam. For spectroscopy of the $^1S_0 - {}^3P_0$ transition, the coupling magnetic field of 2.3 mT is turned on and a 200 ms long pulse of the 698 nm interrogation laser with an intensity of 3.2 W/cm$^2$ excites a fraction of the atoms into the $^3P_0$ state. The 461 nm MOT beams are then used to detect the remaining ground state atoms by their fluorescence.

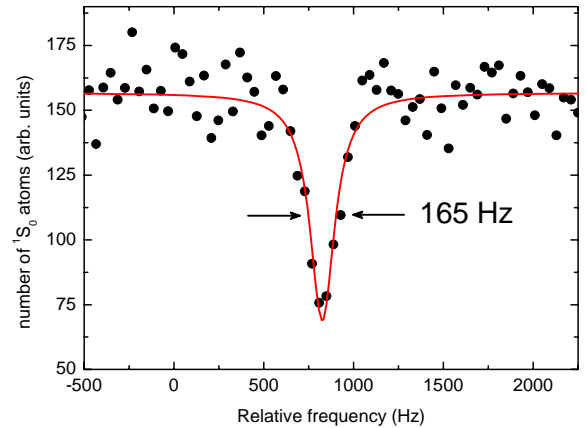

Fig. 5: Magnetically induced $^1S_0 - {}^3P_0$ clock transition of $^{88}$Sr. Each data point corresponds to a single measurement cycle of 640 ms. The frequency axis is determined from the offset between the reference cavity and the interrogation laser (twice the AOM frequency). It is corrected for the cavity drift and shifted by an arbitrary frequency to obtain small numbers.

Fig. 5 shows the variation of the fluorescence signal with respect to the interrogation laser frequency. The cycling time of the frequency scan is given by the duration of the cooling stages, the probe pulse and the fluorescence detection, which sums up to 640 ms.

With a Lorentzian line fit to the measured spectrum we get a FWHM linewidth of 165 Hz. The difference to the laser linewidth deduced from the virtual beat is most likely due to different environmental conditions, e.g. no acoustic isolation box was used during the spectroscopy. To achieve better resolution in future measurements, it is necessary to improve the probe laser linewidth. This is achieved by phase locking the laser to the 657 nm reference laser as described in the next section.

## V. PHASE LOCK

To perform a phase lock of the 698 nm diode laser system to the 657 nm reference laser, the virtual beat signal between both lasers is compared with the output of an rf-synthesizer by a phase and frequency comparator ($\Phi$, Fig. 6). The rf-synthesizer is referenced to a 100 MHz signal derived from a H-Maser. The comparator output drives a high quality surface-acoustic wave (SAW) 400 MHz voltage controlled oscillator (VCO). A direct digital synthesizer (DDS) is used to transform this signal to the frequency of 266 MHz driving the double-pass acousto-optical modulator (AOM) between the Sr master laser and the reference cavity. The rf-frequency at the AOM controls the Sr laser frequency and therefore the virtual beat frequency, thus closing the phase-locked loop.

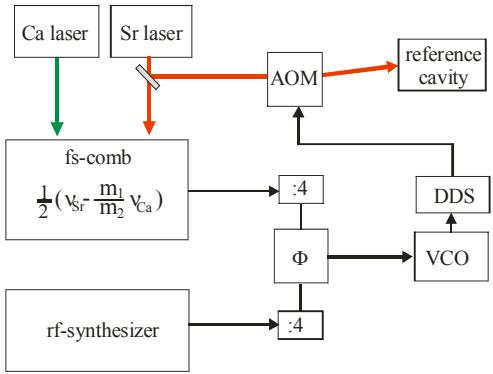

Fig. 6: Schematic of the phase lock of the 698 nm Sr laser to the 657 nm Ca laser serving as a reference.

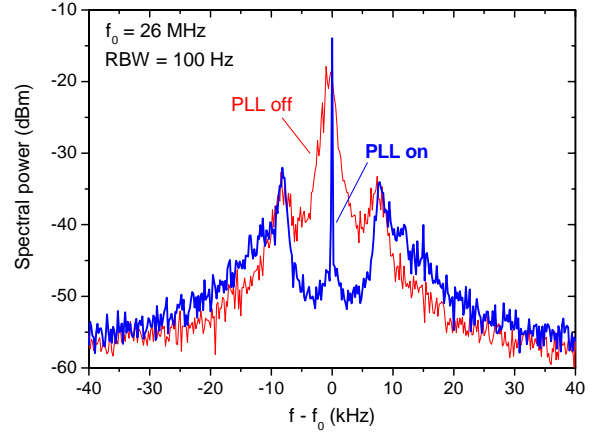

Fig. 7: Virtual beat between Sr and Ca laser with the phase-locked loop (PLL) open and closed. With the PLL open the noise of the free running VCO is dominating. Both spectra are averaged over 10 sweeps with a sweep time of 80 ms.

Fig. 7 shows the spectrum of the virtual beat at 26 MHz with the PLL of the 698 nm Sr laser open and closed, respectively. Without the PLL the laser showed a linewidth of a few hundred Hertz caused by seismic and acoustic vibrations of the reference cavity; with open PLL we observe frequency noise of the free-running VCO. When the phase-locked loop is closed, the virtual beat signal narrows down to a $\delta$-function. The measured linewidth of the virtual beat of 1 Hz is then limited by the resolution bandwidth of the spectrum analyzer. The virtual beat signal between the phase-locked lasers is an in-loop signal and does not show possible additional phase noise from the fs fiber comb, the electronic components, and uncompensated fiber links. Previous tests of the transfer method using two independent combs have confirmed that a linewidth of less than one Hertz can be achieved [3]. Further spectroscopy on the Sr clock transition including fiber noise cancellation [16] could be used to test the laser linewidth. So far, different technical problems concerning the experimental setup have not allowed spectroscopy on the clock transition using the phase locked laser.

## VI. CONCLUSIONS

We have set up a 698 nm laser system to investigate the clock transition of bosonic $^{88}$Sr induced by a magnetic field. The laser was characterized relative to an ultranarrow linewidth reference laser at 657 nm using a femtosecond fiber comb as transfer oscillator. The comparison is unaffected by fluctuations of the repetition rate and the carrier envelope frequency of the comb and provides a real time virtual beat between the two lasers with large frequency bandwidth. As the measured linewidth of the clock transition had been limited by the interrogation laser linewidth, we have narrowed the laser linewidth by phase-locking the virtual beat to a stable rf-reference. With this method the stability and linewidth of a given reference laser can be transferred to a second laser at an arbitrary wavelength within the spectral range accessible by the frequency comb.